# Brainy light sensors with no diffraction limitations

**H.J. Caulfield[1], L. P. Yaroslavsky[2], Jacques Ludman[3]**

[1] *Physics Department, Fisk University*
*1000 17th St., Nashville, TN 37298, USA*
[2] *Dept. Interdisciplinary Studies. Faculty of Engineering. Tel Aviv University.*
*Tel Aviv, Ramat Aviv 69978, Israel*
*3 Jacques Ludman*
*Northeast Photosciences, Inc.*
*18 Flagg Road, Hollis, New Hampshire 03049*

**Abstract:** In total ignorance of what a scene contains, imaging systems are extremely useful. But if we know the scene will be comprised of no more than a few distant point sources, nonimaging systems may achieve better accuracy in a smaller, more rugged, and less expensive manner. We show here that those advantages can be realized in a wide variety of designs. All can beat the diffraction limit under the proper circumstances. We call these sensors "brainy" in analogy to anima; vision which uses poor optics processed by a wonderful computer – a brain.

## References and links

## Introduction

A simple plane wave incident on a sensor has only three parameters: two directions of arrival and an intensity. If it is known *a priori* that there is only one such beam, extracting those three parameters should require only three measurements and simple mathematical inversion.

Localization of light sources is a decision-making process. A lens does this quite well. It processes, in parallel, the incident wavefront and translates the two angles of the incident beam into two spatial coordinates in the image plane, and the intensity of the focused point gives information about the intensity of the plane wave. As a result, decision-making and estimation becomes trivial and do not require any non-point-wise operations. Thanks to the lens, the computational complexity of localization is proportional to the number of points to be resolved in the image. The price of this solution is diffraction limitation.

The role of the lens in the case of a single distant point source is to perform an optical Fourier transform on the incident light, thus concentrating the light from the point source but not concentrating the noise. That task could be performed by measurement in the incident plane and numerical Fourier transformation, but use of the lens accomplishes the task optically prior to detection. This makes the point location very simple electronically. It reduces the computation in proportion to the computational complexity of the numerical Fourier transform that would have been used without the lens. The lens does play the role of a matched filter which maximizes signal-to-noise ratio at its output, but with respect to noise in the incident beam, and not with respect to noise of sensors placed in lens' focal plane. In the application discussed here, this property of the lens is irrelevant.

Such considerations might cause one to wonder why anyone would use anything other than a lens for this problem.. Imaging systems for distant point location and intensity measurement must be judged by such figures of merit as

- Accuracy
- Resolution (a very closely related consideration)
- Cost
- Physical space consumed
- Robustness

Nonimaging systems may offer advantages in those areas.

One way to view the situation is that we can do a Bayesian inversion from the measured data to the sought after conditions of angles and intensity. So any way of detecting signals that depend on those unknowns becomes a possible measurement approach.

Thus, non imaging systems can be explored when we know *a priori* that the scene is composed of one or even a few point sources at infinity. Bayes's theorem teaches us how to combine current measurements with prior knowledge in the presence of sensor's noise to obtain the intensities and locations of those. For point sources of light at infinity, image forming is unnecessary. All we require is those three data for each source.

To our knowledge, these sorts of approaches have never been considered in any detail. Nonimaging point location is certainly not a new concept. Holography and its noncoherent cousin Coded Aperture Imaging systems are examples as are the various forms of medical tomography. These fields will be well known to most readers of this work. For example, we are among the many who have published books in this field [1-4]

The closest we have found to our work is the work of Luo et al [5]. Particles are treated as point sources and the light from them is sampled by nonimaging concentrators. So, in spirit, this is related closely**,** but it is quite different. By using readily available flat detector arrays, the brainy sensors become less expensive, and more plentiful.

There are obvious potential advantages to nonimaging interrogation of the incoming light.. For example, we can avoid the expense and fragility of high quality lenses

As lenses require space to effect the transformation to an image, they require a considerable volume of space. But the various nonimaging systems we explore here do not use image formation and can thus be much smaller and much more robust.

As no imaging is involved, diffraction limits of imaging systems have no direct bearing on achievable resolution.

The essence of what Rev. Bayes taught us is that if we have *a priori* information or physical measurements, it is foolish not to use that information in solving the inverse problem of determining as best we can the values of angular location and intensity from any set of measurements. For this, we should supplement sensors with a “brain” for signal processing and decision making based of sensor output signals.

Any computational transformation of the detected image data is called image processing. If we do not actually compute an image but compute instead the parameters that characterize an image, the possibility of achieving the listed advantages appears.

There are many systems that use rather poor imaging systems and still achieve good results by learning to process the received data extremely well. Those are the visual systems of most animals. Brainy sensors are much cheaper and more robust than perfect optical sensors. We explore here the design of brainy point source sensors for technological rather than biological uses.

Applications are many. For example, the military likes to obtain that location information about incoming objects that are too small to image. Space ships and other systems require star trackers to navigate. Locating sources of laser light illuminating me is of interest in some cases. Locating and tracking LEDs built into or attached to physical structures (bridges or aircraft for example) allows us to verify their behavior under load and to test for inelastic events that may indicate ageing. Radioactive source location is important in many field. As all we require is a suitable detector array, Robots may be asked to go to places specified by beacons at different locations and colors.

In the balance of this paper, we will analyze some of these simple nonimaging point locators in some detail before seeking generalizations supported by those analyses.

## 1. Formulation of the problem and an analytical solution

Consider the configuration of point parameters shown in Fig.1. This defines the parameters of the physical; setup for a single point source.

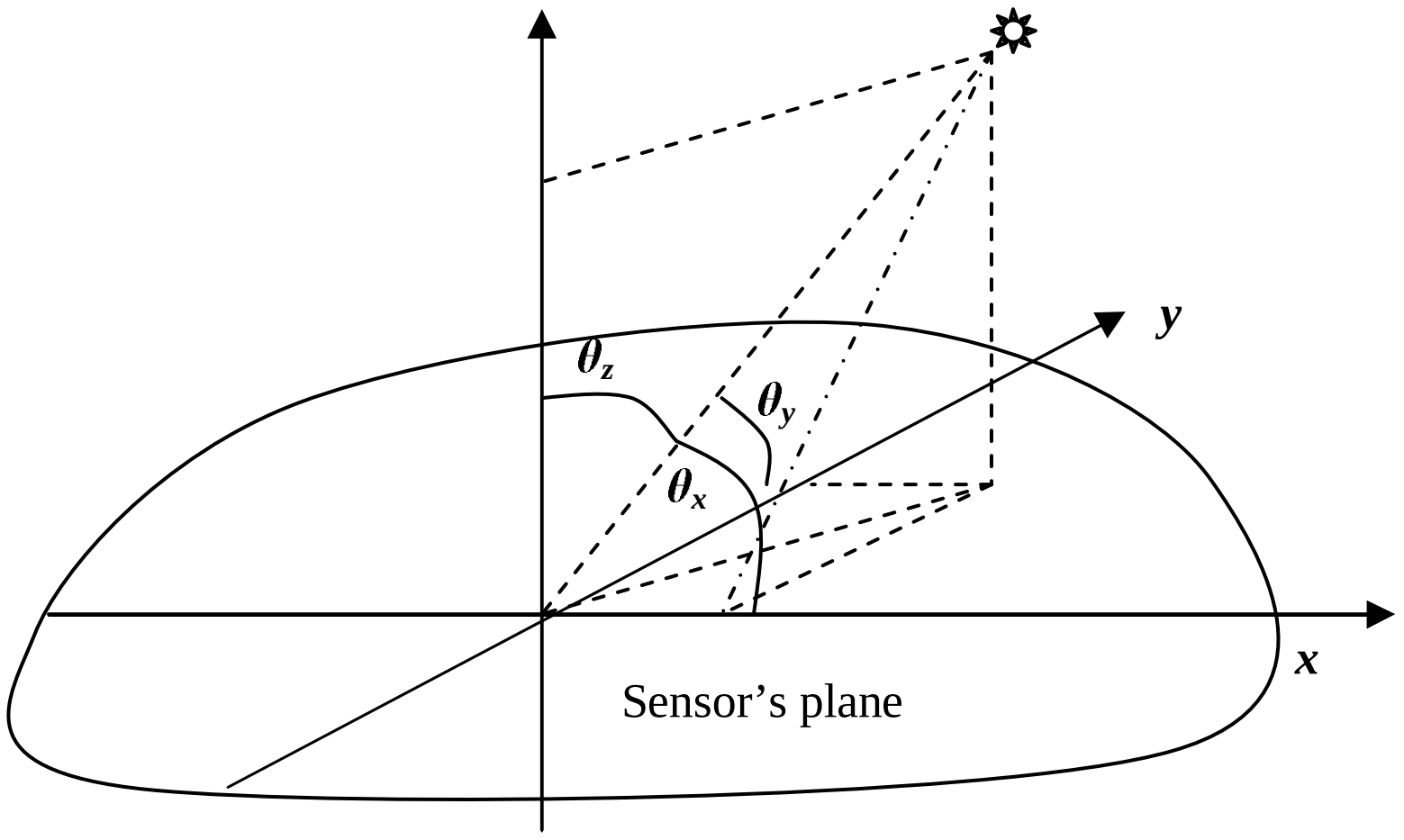

Fig. 1. The point location relative to the center of the observation plane is defined here.

The purpose is determination of intensity and directional angles $\{\theta_x, \theta_y\}$, in the sensor’s coordinate system, of a single source of light. Estimation of these three parameters requires at least three “independent” measurements. The word “independent” is put in quotation marks, because although independence is desirable, all we really need is three measurements each of which depends on the sought after parameters in a known way.

We consider using plane sensors with “natural” directivity: sensor’s response to illumination is proportional to cosine of the angle between the direction to the source and normal to the sensor’s plane. That is, the suffer from obl;iquity effects that have been known for centuries. We will also assume that sensor’s signal contains, in addition to the proper signal, additive signal independent normal sensor’s noise. It suffices that the measurements be invertible by Bayesian methods and independent of one anther.

A single flat detector element has an angular dependence because of the obliquity effect. That sensitivity can be modified by tiling the plane with detector structures that deviate from local flatness as suggested in Fig. 2. In this case, the peak sensitivity directions are the normals to the detectors. Note that this configuration contains a simple flat array as a special case. .For simplicity, in what follows we consider the plane (2D) model shown in Fig,2.

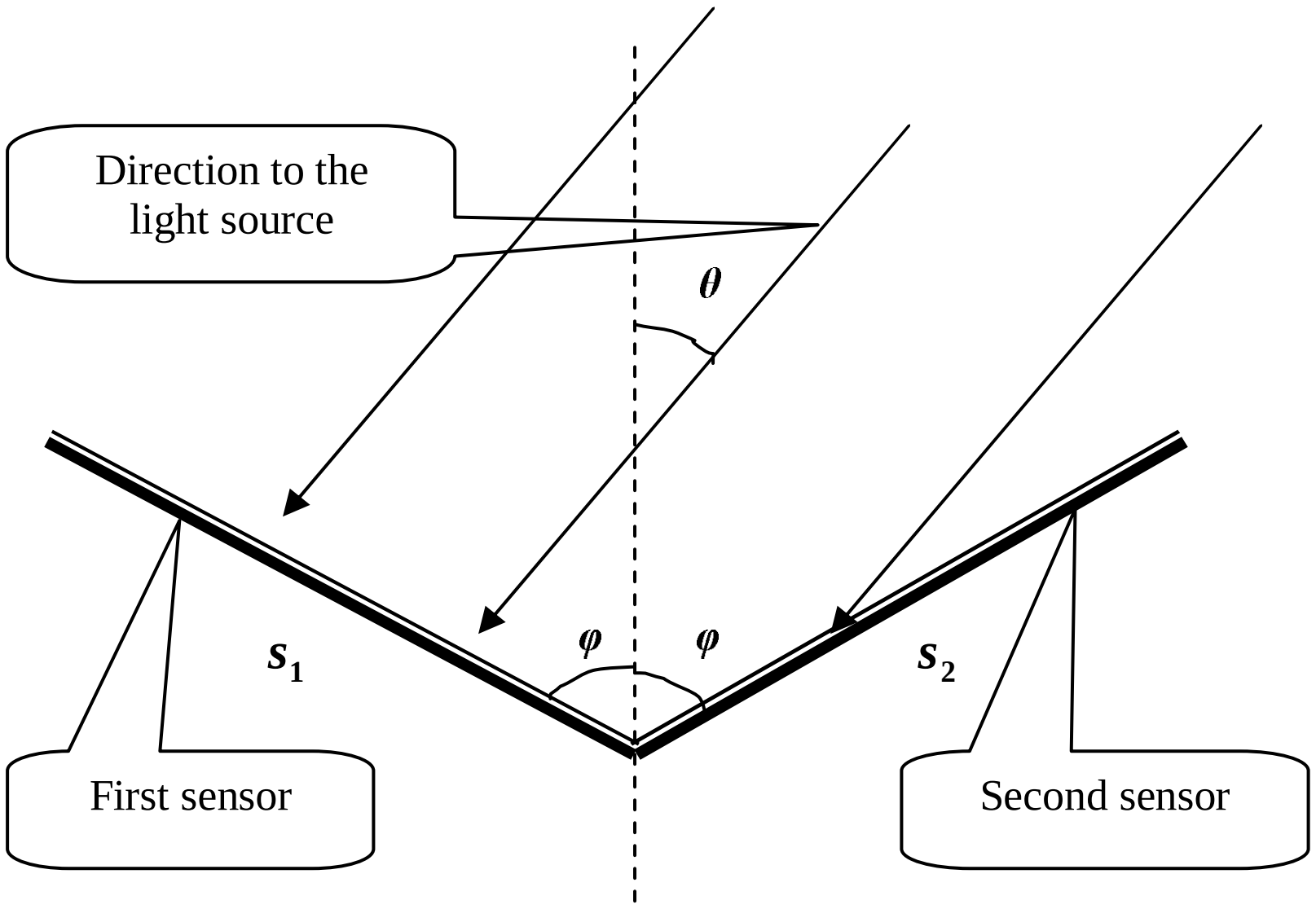

Fig. 2. Two differently oriented plane sensors with cosine-law directivity for determination of the directional angle and intensity of the light source.

In 2D case, three parameters, two directional angles and intensity of the light source must be determined. Shown here two differently oriented plane sensors with cosine-law directivity are sufficient.

Let $A$ and $\theta$ be intensity and direction angle of a point source, $s_1$ and $s_2$ are sensor’s output voltages. In the assumption of signal independent additive normal sensor’s noise, maximum likelihood estimations $\left(\hat{A},\hat{\theta}\right)$ of the source intensity and directivity angle are defined by equations:

$$\left(\hat{A},\hat{\theta}\right)=\arg\min_{\hat{A},\hat{\theta}}\left\{\left[s_1-\hat{A}\sin(\varphi+\hat{\theta})\right]^2+\left[s_2-\hat{A}\sin(\varphi-\hat{\theta})\right]^2\right\}=$$

$$\arg\min_{\hat{A},\hat{\theta}}\left\{\left[s_1-\hat{A}_c\sin\varphi-\hat{A}_s\cos\varphi\right]^2+\left[s_2-\hat{A}_c\sin\varphi+\hat{A}_s\cos\varphi\right]^2\right\} \qquad (1)$$

where $\hat{A}_c = \hat{A}\cos\hat{\theta}$; $\hat{A}_s = \hat{A}\sin\hat{\theta}$.

These equations have the following solutions:

$$\hat{A}_c = \frac{s_1 + s_2}{2\sin\varphi};\ \hat{A}_s = \frac{s_1 - s_2}{2\cos\varphi};\ \hat{A} = \frac{\sqrt{s_1^2 + s_2^2 + 2s_1 s_2 \cos 2\varphi}}{\sin 2\varphi} \tag{2}$$

$$\tan\hat{\theta} = \frac{\hat{A}_s}{\hat{A}_c} = \frac{s_1 - s_2}{s_1 + s_2}\tan\varphi, \tag{3}$$

In the assumption of high signal-to noise ratio, one can also obtain that estimation errors $\{\varepsilon_A, \varepsilon_\theta\}$ due to the sensor noise have normal distribution with zero mean and variances, correspondingly:

$$\overline{\varepsilon_A^2} = \frac{1 + \cos 2\theta \cos 2\varphi}{\sin^2 2\varphi}\sigma_n^2 \tag{4}$$

and

$$\overline{\varepsilon_\theta^2} = \frac{1 - \cos 2\theta \cos 2\varphi}{\sin^2 2\varphi}\frac{\sigma_n^2}{A^2}, \tag{5}$$

where $\sigma_n^2$ is variance of the sensor's noise.

For $\varphi = \pi/4$, error variances are minimal and do not depend on the source directional angle:

$$\overline{\varepsilon_A^2} = \sigma_n^2;\ \overline{\varepsilon_\theta^2} = \frac{\sigma_n^2}{A^2}; \tag{6}$$

Variance of the angle estimation error determines accuracy of the source directional angle measurement and, therefore, sensor's resolving power. A remarkable property of such a sensor is that its resolving power has no diffraction limitations. It depends only on signal-to-noise ratio $\sigma_n^2 / A^2$, which, in its turn, is determined by the sensor's size (area), whereas the resolving power of conventional sensors placed in the focal plane of the lens, is determined, in addition to signal-to-noise ratio, by diffraction limitations that exhibit themselves through the size of the lens.

## 2. Other Possible Designs

In order to make the described sensor as flat as possible and to increase its field of view, the following two designs can be suggested.

Recently, we have begun to explore another means for determination of the direction angle and intensity of a single source of light, a plane sensor that utilizes widely available plane photosensitive arrays. In this sensor, plane photosensitive array is placed behind a lens (Fig. 4, a). In this sensor, lens plays a role of a set of prisms (Fig. 4, b)) that change direction of light that arrives at a particular elementary sensor and, therefore, converts the plane array of sensors with identically oriented elementary sensors into a set differently oriented sensors, which is required for enabling localization of a point source of light according to above described principle. The hope was that using this approach would allow the use of readily available detector arrays rather than make new ones.

Extending this concept from 1D to 2D is straightforward and will not be explored here. One can also consider an option of placing sensor cells on a curved surface as it is illustrated in Fig. 5. In this case, no optics is needed at all.

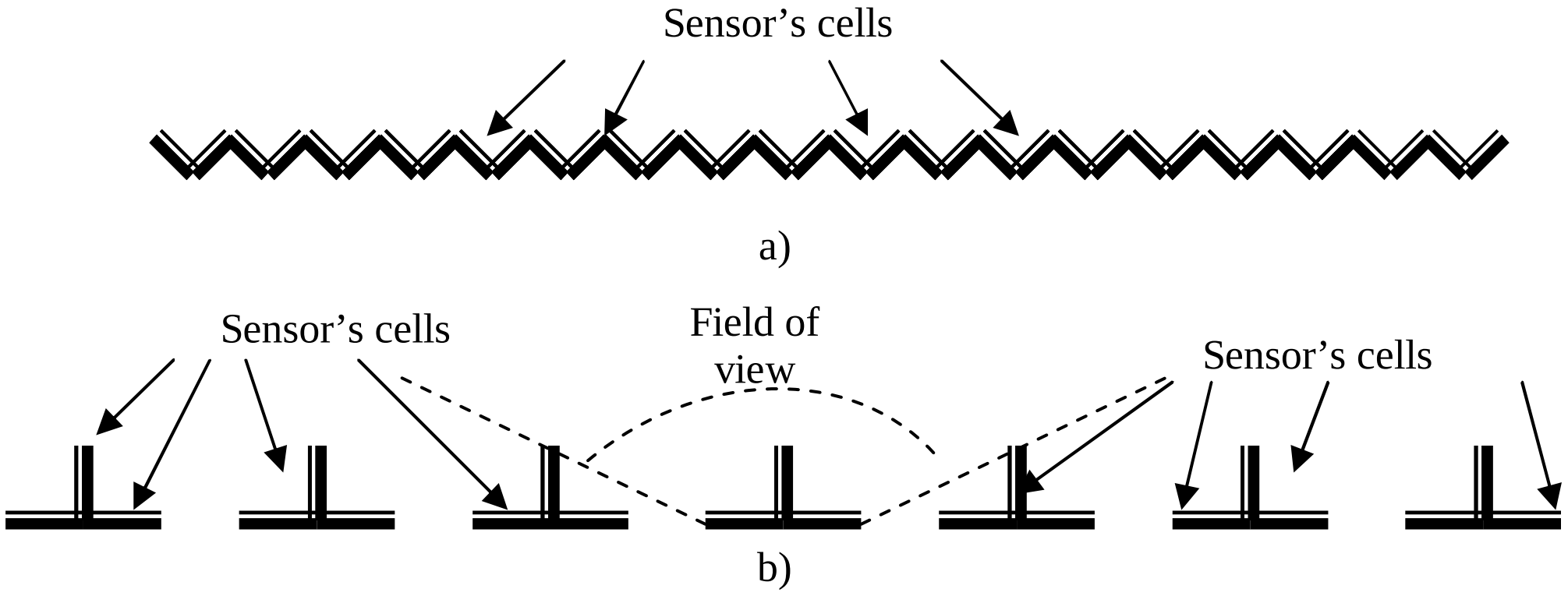

Fig. 3. Two possible sensor array designs from multiple small elementary sensors

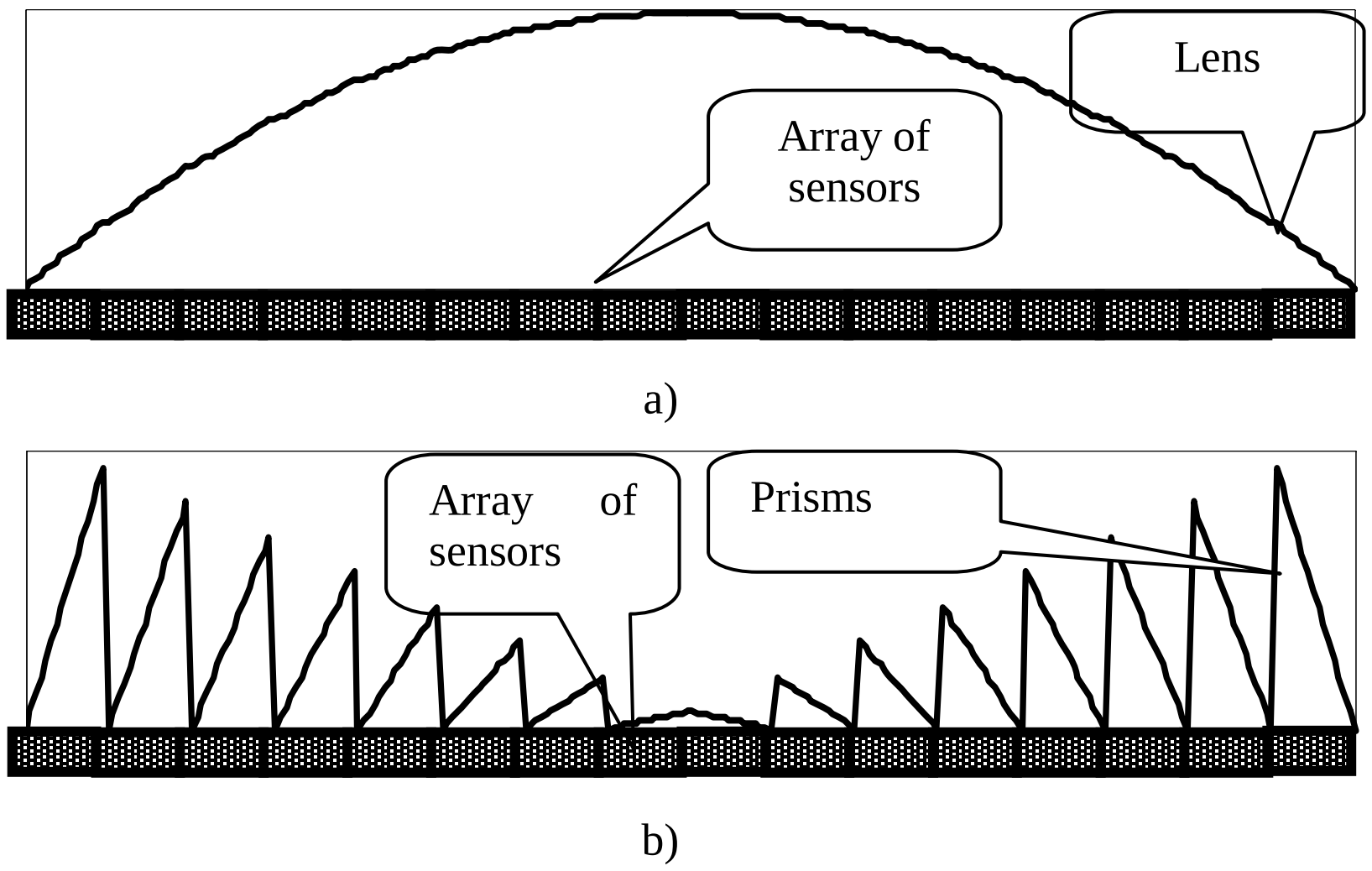

Fig. 4. The plane sensor (a) and its equivalent representation (b) .

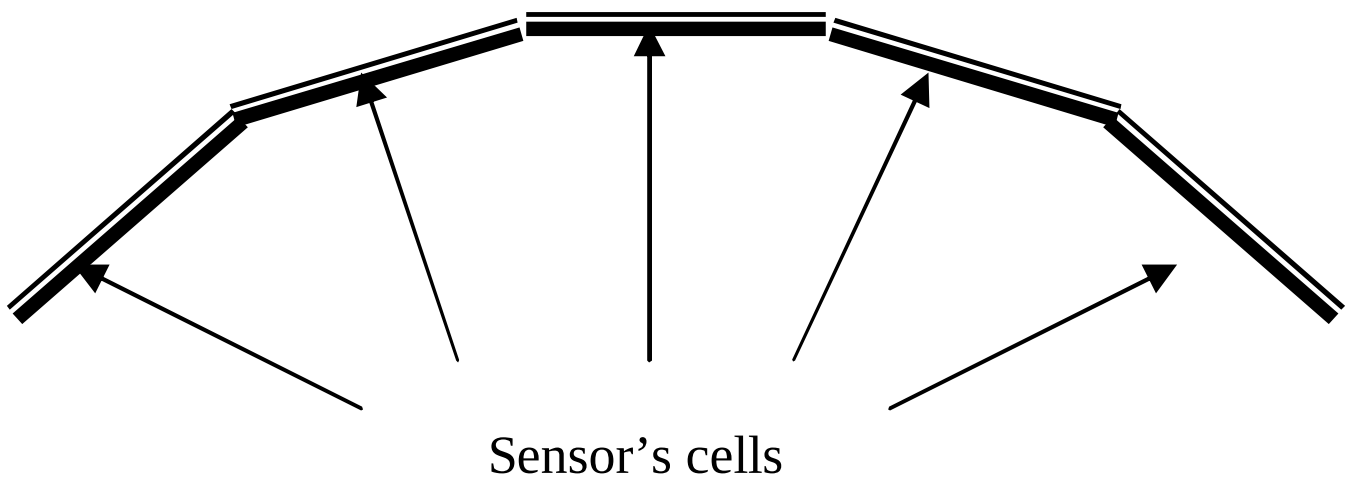

Fig. 5. The array of multiple cell "plane" lens free sensor placed on a curved surface.

That is, we can detect a flat, plane wave with sensor elements on a curved surface (Fig. 5) or detect a curved wavefront on a flat sensor array. The use of a lens makes the system considerably less expensive, but the same theory underlies both approaches.

Some of these systems may not have convenient closed-form sensitivity curves. But measurements can calibrate them well enough to make Bayesian inversion reliable and accurate. We have chosen to explore some of these concepts using sensors and other components that can be described easily by equations. The results will be quite general though.

## 3. Simulation results of the “plane” sensor array for localizing a single point source of light

We investigated sensitivity and angular resolving power of the “plane” sensor of Fig. 4 to a point light source using a 2-D Monte-Carlo computer model of the “plane” sensor array with 100 elementary sensors. The sensitivity and angular resolving power of the sensor are characterized by standard deviation of the light intensity and directional angle, correspondingly. Results regarding standard deviations of the intensity and angle estimations obtained are summarized in Figs. 6 and 7

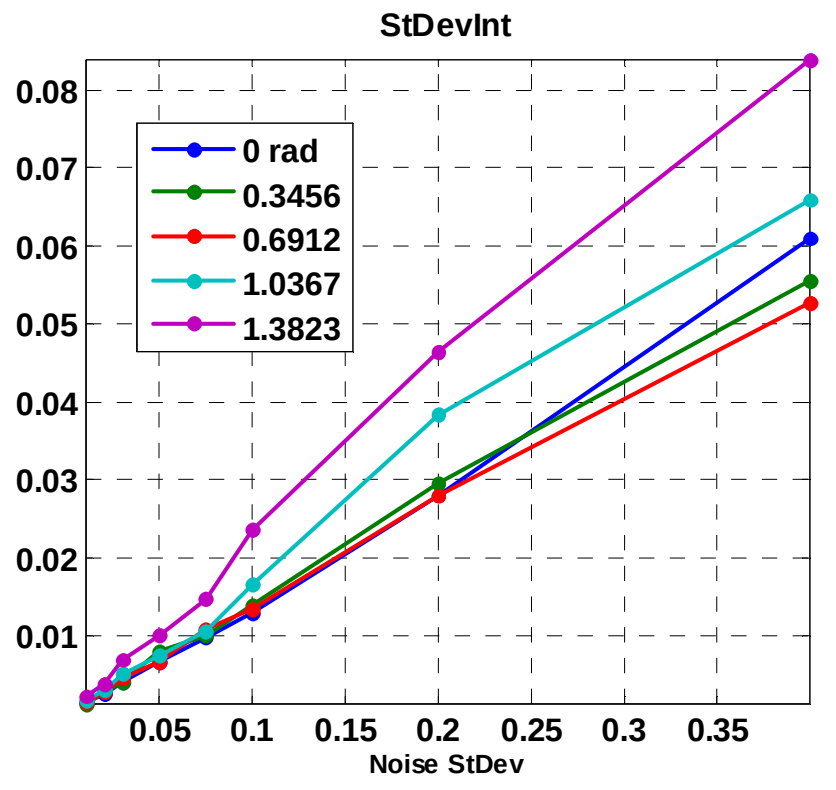

Standard deviation of intensity estimation error vs sensor noise level

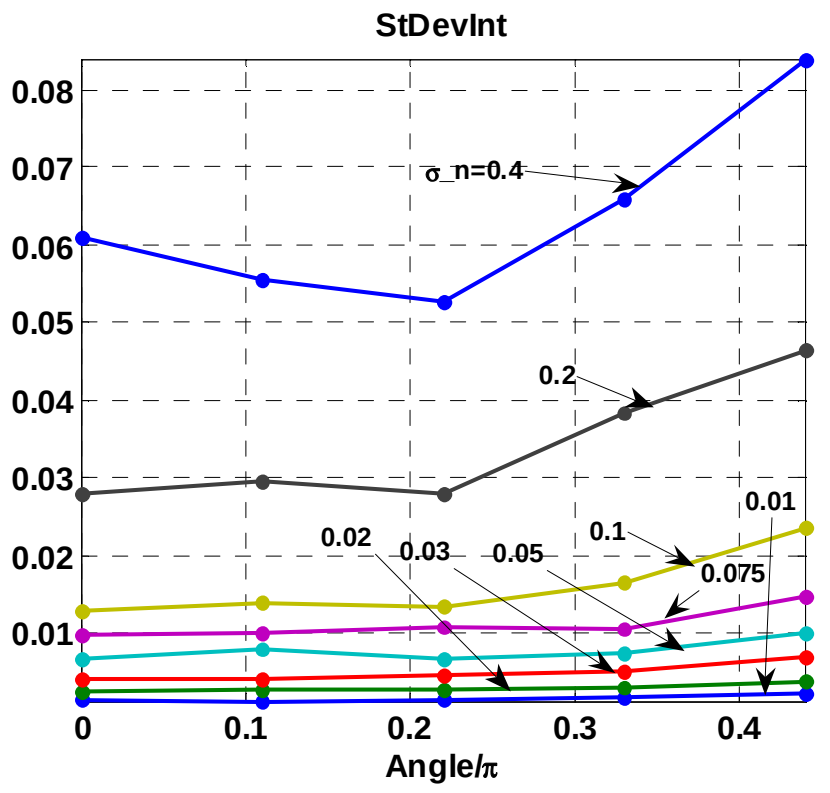

Standard deviation of intensity estimation error vs source directional angle

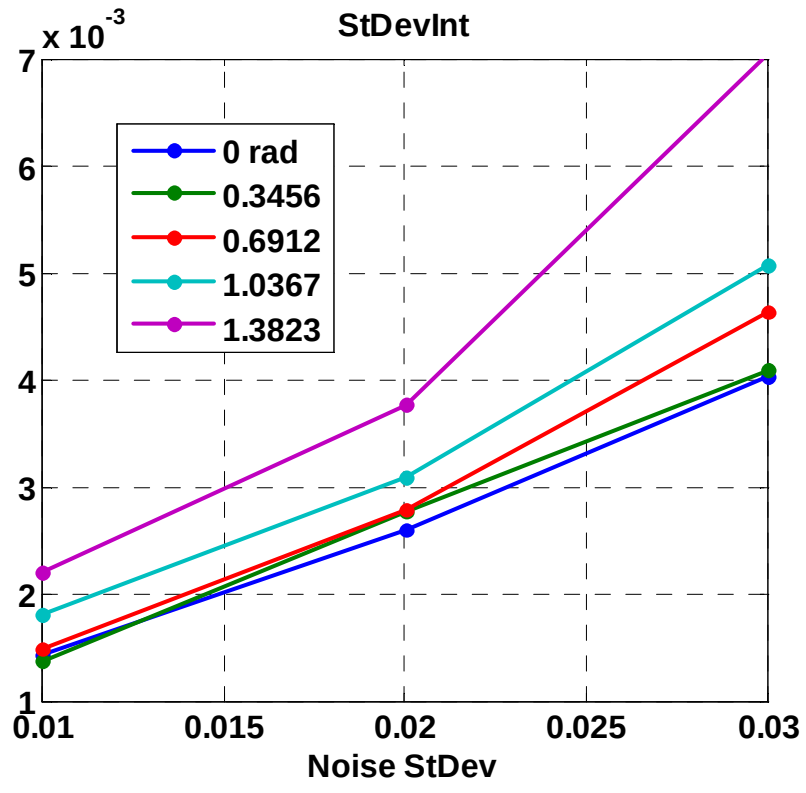

Plots of a sub-set of above data for noise levels between 0.01 and 0.03

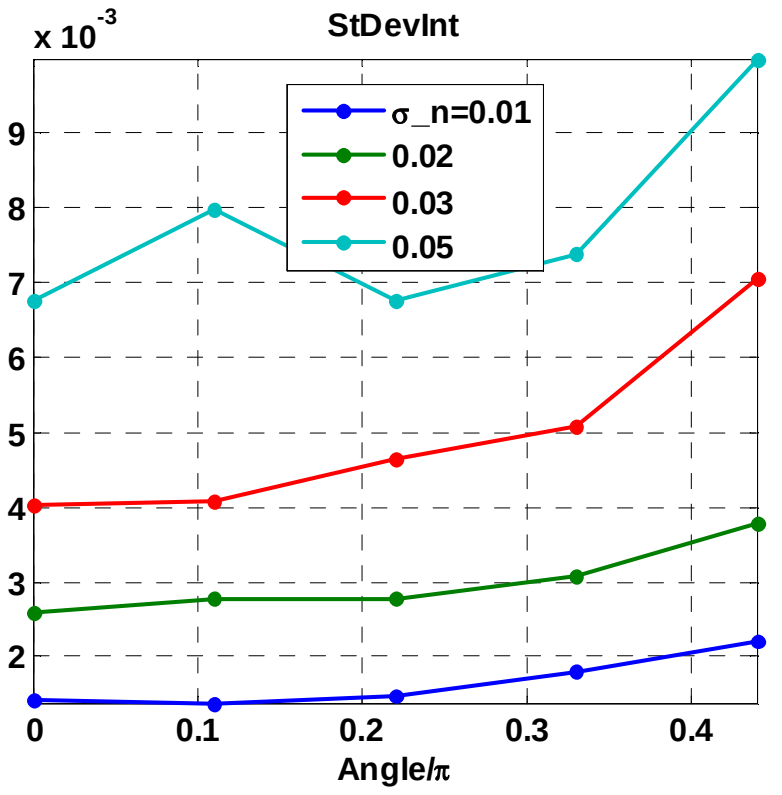

Plots of a sub-set of above data for noise levels between 0.01 and 0.05

Fig. 6. Standard deviation of source intensity estimation error vs. sensor’s noise level (left column) and source’s directional angle (right column).

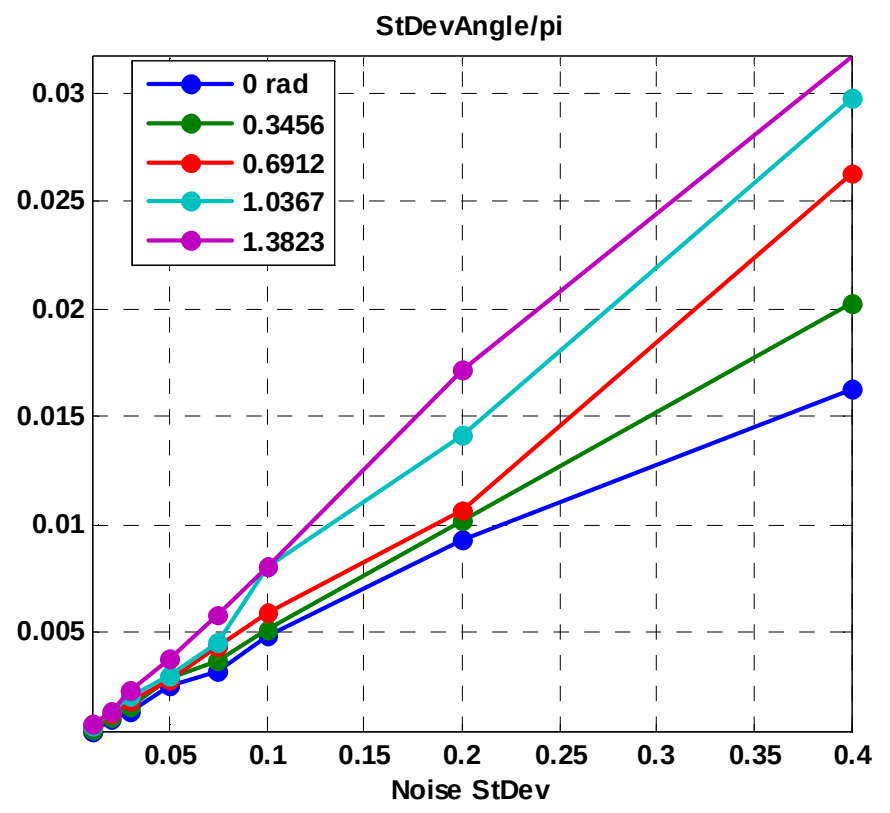

Standard deviation of directional angle estimation error vs sensor's noise level

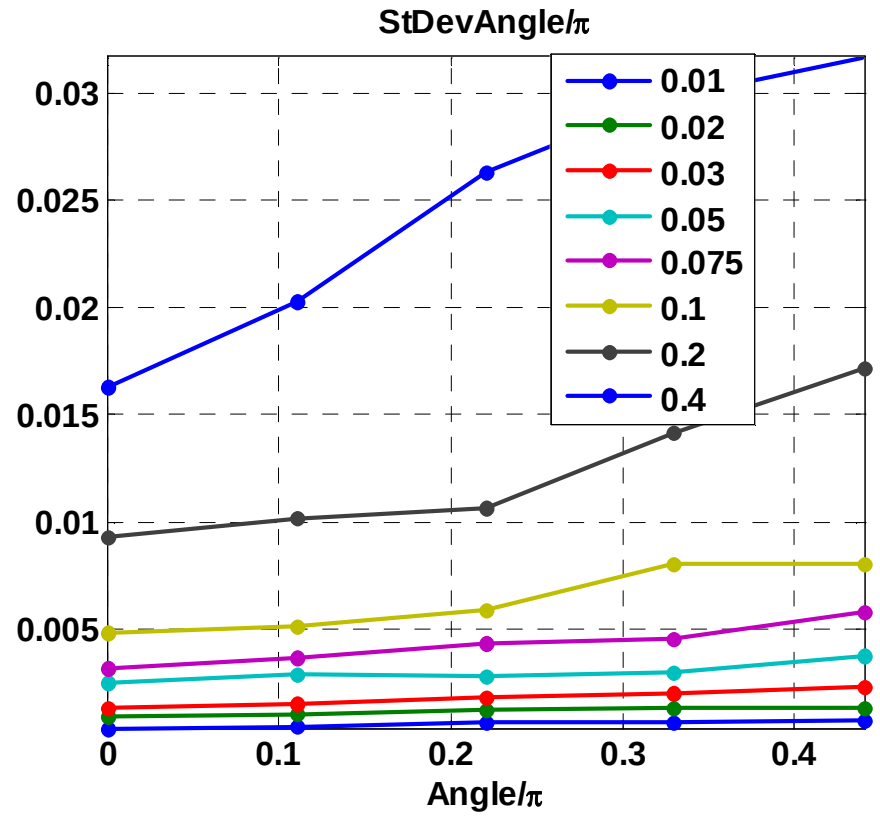

Standard deviation of directional angle estimation error vs source's directional angle

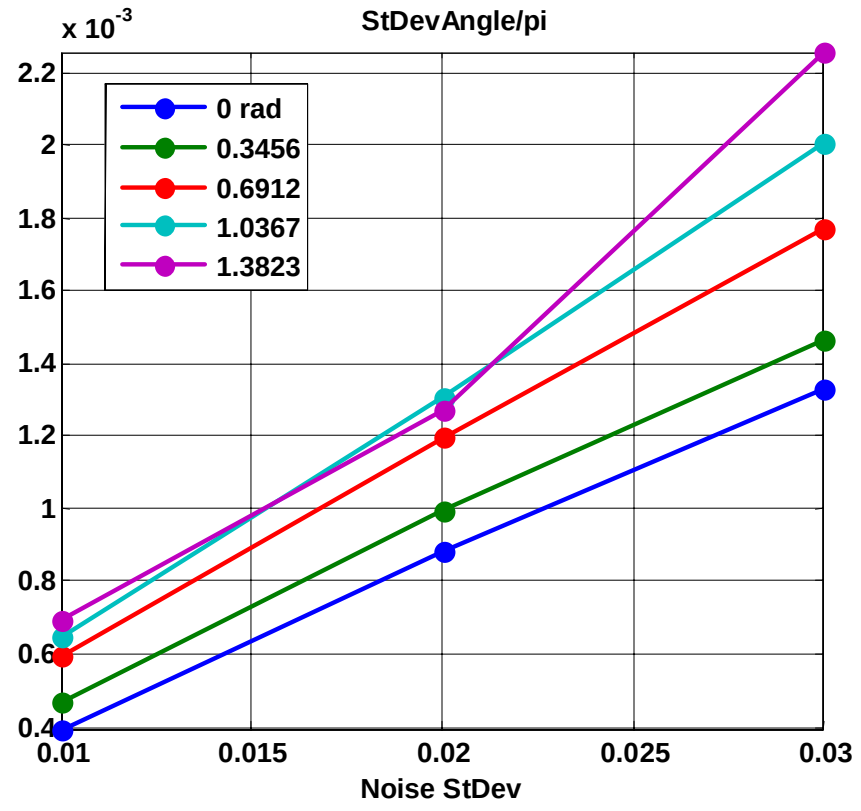

Plots of a sub-set of above data for noise levels between 0.01 and 0.03

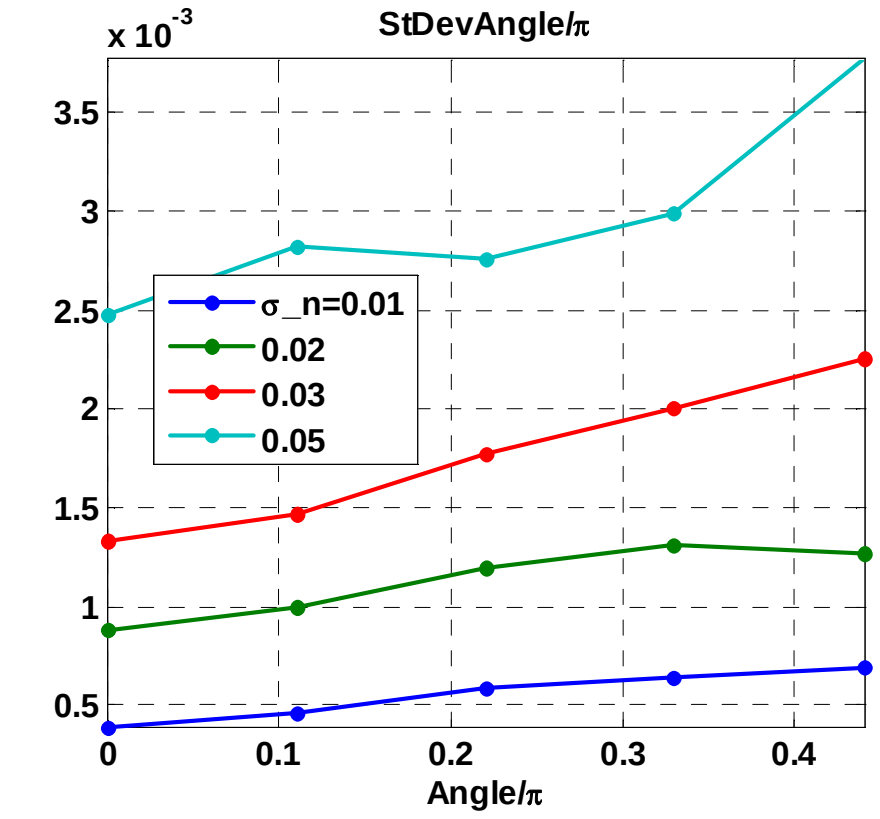

Plots of a sub-set of above data for noise levels between 0.01 and 0.05

Fig. 7. Standard deviation of directional angle estimation error vs sensor's noise level (left column) and source's directional angle (right column).

## 4. Possible extensions

The basic work principle of above-described "plane" sensors is ML-estimation of source intensity and directional angles by means of processing of raw measurements of illumination level of multiple "elementary" differently oriented sensors with "cosine" angular sensitivity. This principle can be in a straightforward way extended to:

- Determination of the intensity, directional angle and the distance to a single point light source. This adds one additional parameter, the distance from one to the other?? to the above described scheme, which requires at least four differently oriented elementary sensors.
- Determination of intensities, directional angles and distances to a known number of multiple point light sources. This will obviously require at least four times as many differently oriented elementary sensors as the number of sources.

- Imaging by means of determination of intensities of a given number of multiple light sources whose positions in space is known.
- Imaging and depth sensing by means of determination of intensities of a given number of multiple light sources whose directional angles are known

## 5. "Brainy" sensor for localization of multiple sources of light

In 2-D formulation, for N sensors oriented under angles $\{\varphi_n\}$, $n = 1,2,...,N$ with respect to the sensor array axis, maximum likelihood ( ML) estimations $\{\hat{A}_k, \hat{\theta}_k\}$ of intensities $\{A_k\}$ and directional angles $\{\theta_k\}$, of the known number K light sources must be obtained as a solution of the equation:

$$\left(\{\hat{A}_k, \hat{\theta}_k\}\right) = \underset{\{\hat{A}_k, \hat{\theta}_k\}}{\arg\min} \left\{ \sum_{n=1}^{N} \left[ s_n - \sum_{k=1}^{K} \hat{A}_k \sin(\varphi_n + \hat{\theta}_k) \right]^2 \right\},\ k = 1,2,...,K \qquad (7)$$

where $\{s_n\}$, $n = 1,...N$ are signals at output of elementary sensors.

One can show that, for small sensor's noise level, estimation errors have normal zero-mean distribution. As in the case of a single source, standard deviation of angle estimation error characterizes angular resolving power of the sensor, and standard deviation of intensity estimations characterizes the sensor sensitivity. From Eq. 7 it follows that, given the number of sensors and sources, "brainy" sensor resolving power and sensitivity depend only on noise level in the elementary sensors.

We present here some numerical simulation results of solving this problem. The results were obtained using a 2-D computer Monte-Carlo model, in which parameter estimation was implemented through a multi-dimensional numerical optimization procedure. Considered was a "brainy" sensor implemented as a set of elementary sensors placed behind a lens (Fig. 4, a) or on a curved surface (Fig. 5). Figs. 8 through 10 illustrate results of localization of tree and five sources by the array of 10, 25, 50 and 100 elementary sensors for sensor's noise standard deviation 0.01 measured in units of source intensity 0-1. The data on the intensity and directional angle estimations obtained over 50 statistical runs are plotted in coordinates defined by intensity and directional angle. The corresponding numerical data for standard deviations (StDev) of the estimation are summarized in Tables 1 and 2. As one would expect, these data show that, given sensor's noise level, estimation error standard deviations that characterize sensor's sensitivity and angular resolving power, increase with number of sources and decrease with the number of elementary sensor cells,

Table 1. Standard deviation of estimations of directional angles and intensities of three sources for different number of elementary sensors.

| Number of sensors | StDev of the angle estimation | | | StDev of estimation of the intensity | | |
|---|---|---|---|---|---|---|
| | Source directional angle | | | Source directional angle | | |
| | - π/5 | 0 | π/5 | - π/5 | 0 | π/5 |
| 10 | 0.15 | 0.077 | 0.135 | 0.26 | 0.14 | 0.28 |
| 25 | 0.00471 | 0.0045 | 0.0049 | 0.019 | 0.034 | 0.027 |
| 50 | 0.00357 | 0.0025 | 0.0036 | 0.015 | 0.022 | 0.016 |
| 100 | 0.00249 | 0.0025 | 0.0026 | 0.014 | 0.019 | 0.013 |

Table 2. Standard deviation of estimations of directional angles and intensities of five sources for different number of elementary sensors

| N of sensors | StDev of estimation of the directional angle | | | | | StDev of estimation of the directional angle | | | | |
|---|---|---|---|---|---|---|---|---|---|---|
| | Source directional angle | | | | | Source directional angle | | | | |
| | - 2π/5 | - 2π/5 | - π/5 | 0 | π/5 | 2π/5 | - π/5 | 0 | π/5 | 2π/5 |
| 10 | 0.079 | 0.079 | 0.078 | 0.09 | 0.074 | 0.12 | 0.28 | 0.27 | 0.31 | 0.1 |
| 25 | 0.016 | 0.016 | 0.008 | 0.011 | 0.006 | 0.04 | 0.036 | 0.028 | 0.038 | 0.034 |
| 50 | 0.009 | 0.009 | 0.005 | 0.007 | 0.005 | 0.023 | 0.021 | 0.027 | 0.022 | 0.022 |

The use of such a "brainy" sensor for localization of multiple sources of light assumes knowledge of the number of light sources. If, in reality, the number of light sources $N_{re}$ is less than the number of sources $N$ assigned to the sensor to search for, sensor's optimization program finds $N$ sources in $N_{re}$ directional angles by splitting $N - N_{re}$ sources to two or several sources with total intensity equal to that of a real source in the given angle. This property allows using the sensor even when exact number of light sources is not known and only their maximal number is known. In this case, it is sufficient to search this maximal number of sources. Payment for this is an increase of the computational burden of the search.

## 6. "Brainy" sensor for localization of multiple sources of light: the localization capability threshold

Above described "plane" sensor of multiple sources is a very nonlinear parameter estimation device From the parameter estimation theory it follows that there exists a threshold of the localization capability of the sensor, under which no localization is possible ([3,4]). Experiments confirmed that, when the number of elementary sensors in the sensor array is too small for the given number of sources and noise level in elementary sensors, estimations become non-reliable with high probability of anomalous large estimation errors.

Given the number of light sources, this threshold can be described either in terms of the minimal number of required elementary sensors for the given noise level in elementary sensors or maximal allowed noise level given the number of elementary sensors.

This phenomenon of non–reliable estimations is illustrated in Fig. 10, a, where results of estimation of intensity and directional angles of 5 sources using only 10 elementary sensors are shown, and in Fig. 10, b, where distribution histograms of the directivity angles of the sources are presented.

## 7. "Brainy" image sensor

One can assume that directional angles of a given number of sources are known. For instance one can try to evaluate intensity of a given number of sensors uniformly distributed in a given angle of view. In such an assumption, the array of differently oriented elementary sensors with natural angle sensitivity can be used as an image sensor. In this case, Eq. 7 for ML estimations of source intensities is reduced to

$$\left(\left\{\hat{A}_k\right\}\right) = \arg\min_{\left\{\hat{A}_k\right\}} \left\{ \sum_{n=1}^{N} \left[ s_n - \sum_{k=1}^{K} \hat{A}_k \sin(\varphi_n + \theta_k) \right]^2 \right\}, \; k = 1,2,...,K \qquad (7)$$

where $\{s_n\}$, $n = 1,...N$ are signals at output of elementary sensors and $\{\theta_k\}$ are known directional angles of the sources.

Fig. 11 illustrates simulation results of 100 runs of sensing, using “brainy” sensor with 100 elementary sensors, of 17 sources of light with known directional angles.

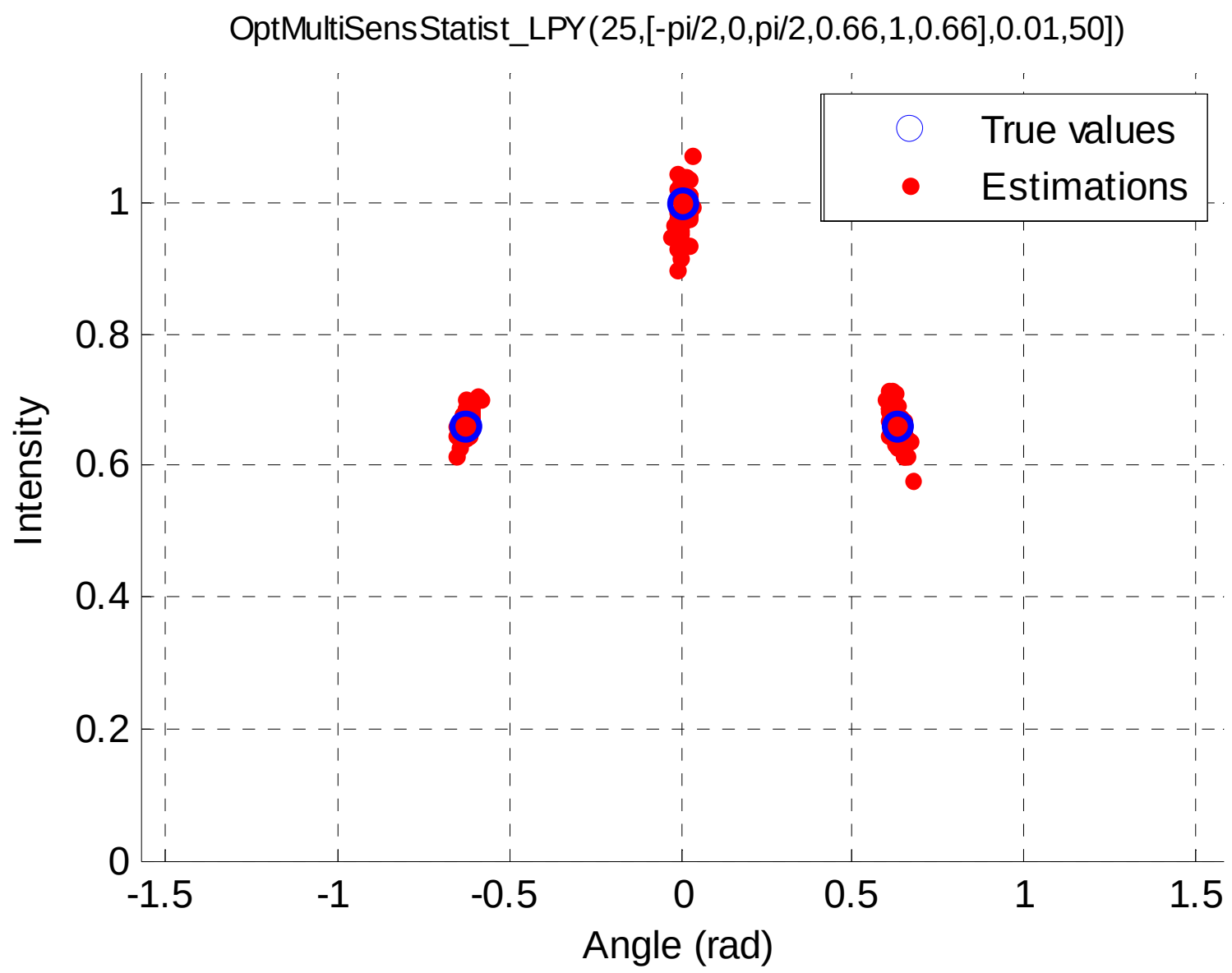

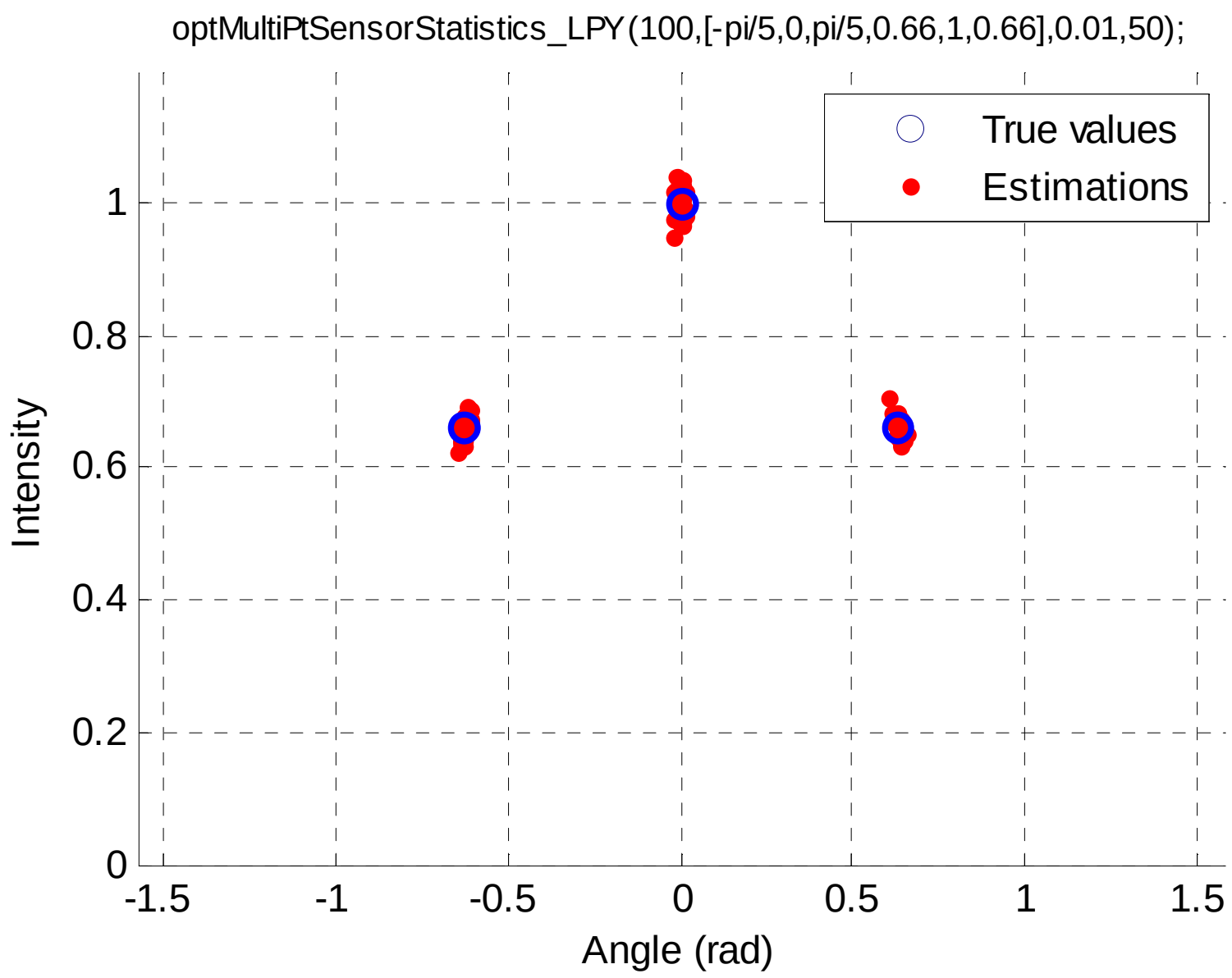

Fig. 8 Three light sources, 25 (upper plot) and 100 (bottom plot) elementary sensors, noise standard deviation 0.001.

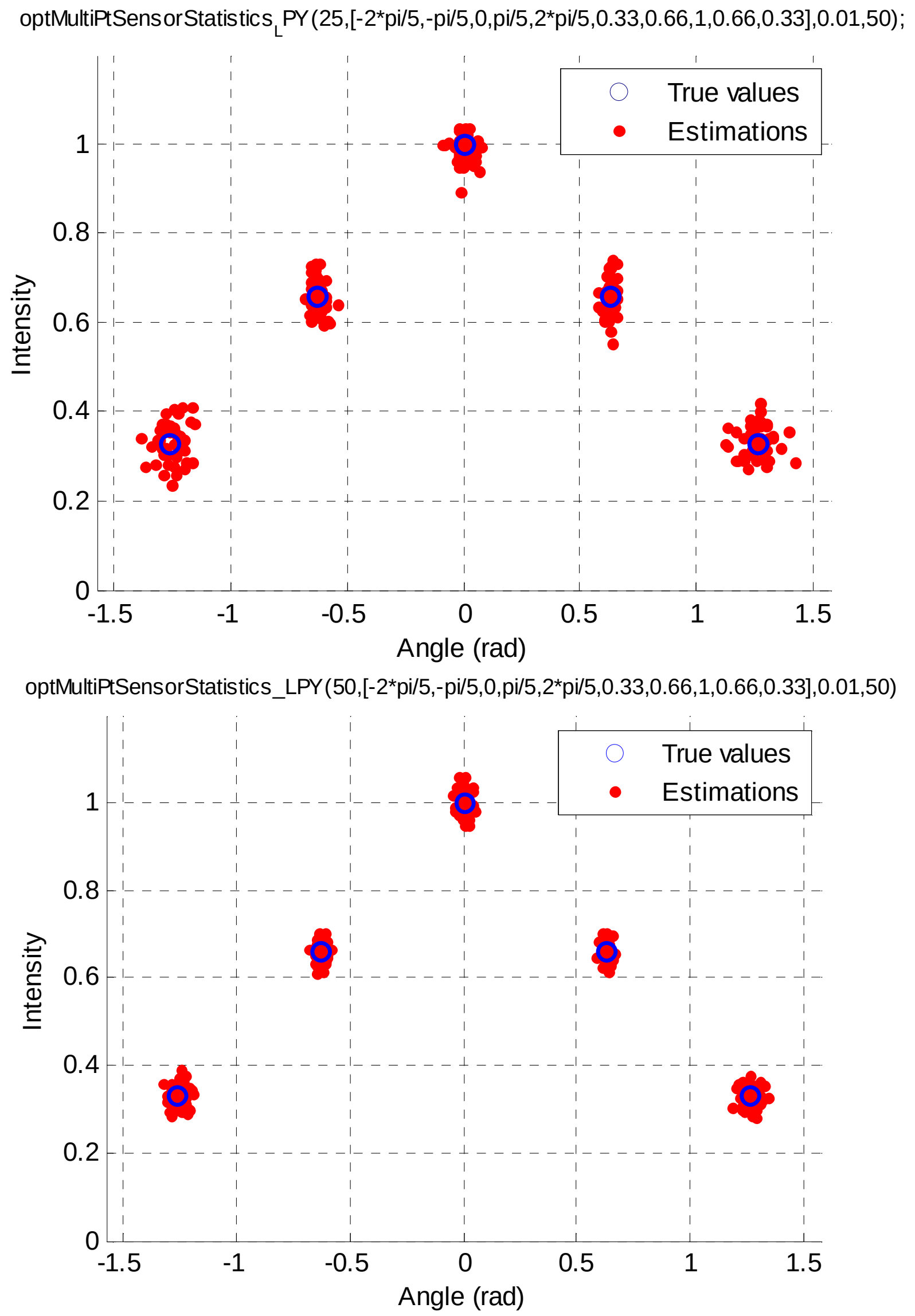

Fig. 9. Five light sources, 25 and 50 elementary sensors, noise standard deviation 0.001 (for the source intensity range 0-1). Compare spread of the results with that in Fig. 8

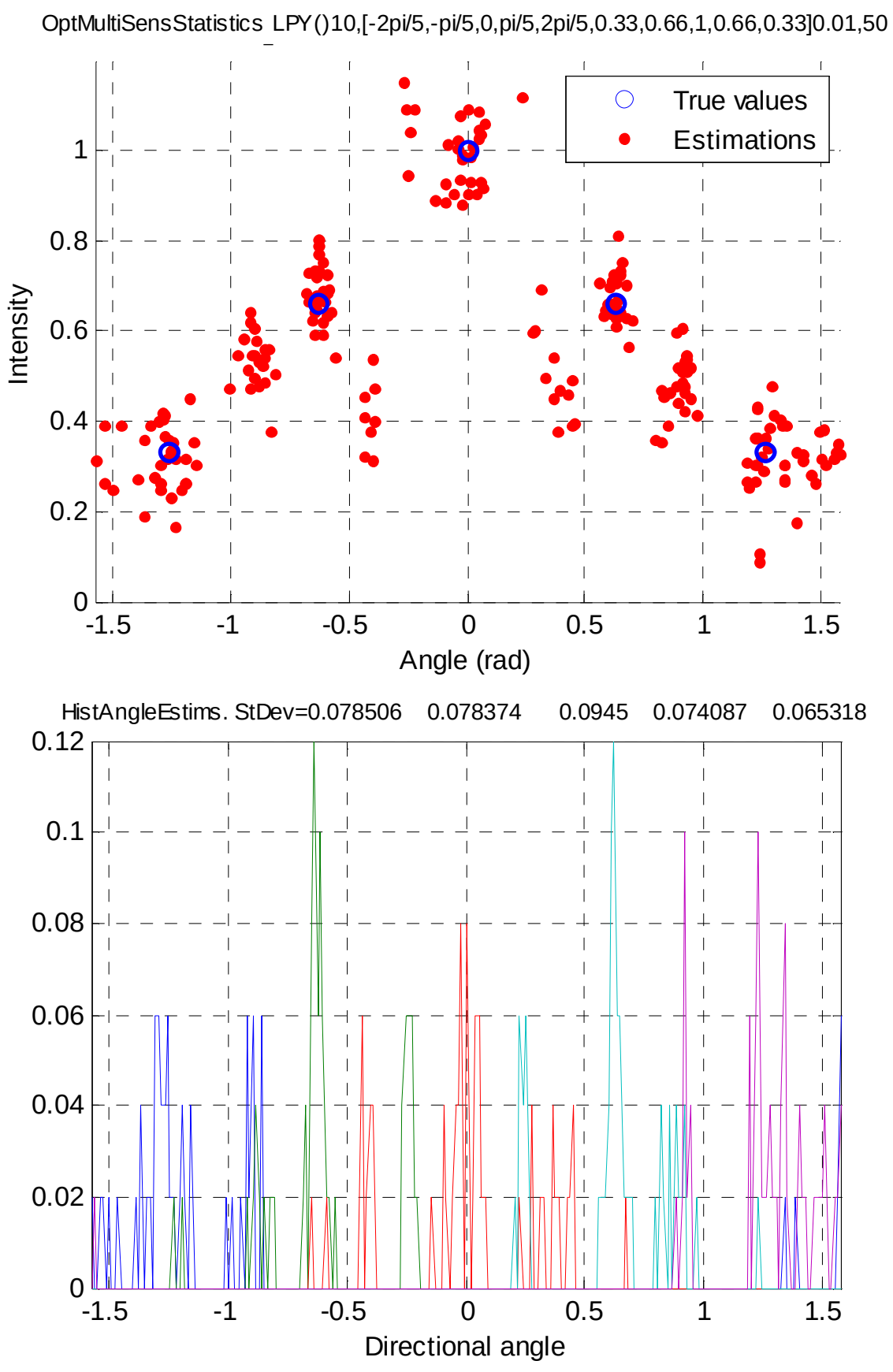

a)

b)

Fig. 10. Diagram of the distribution of estimations of intensity and directional angles of 5 sources (a) and distribution histograms of estimation of their directional angles (b). One can see peaks of histograms that correspond to true source positions and histogram false peaks and long tales due to many cases of false detection

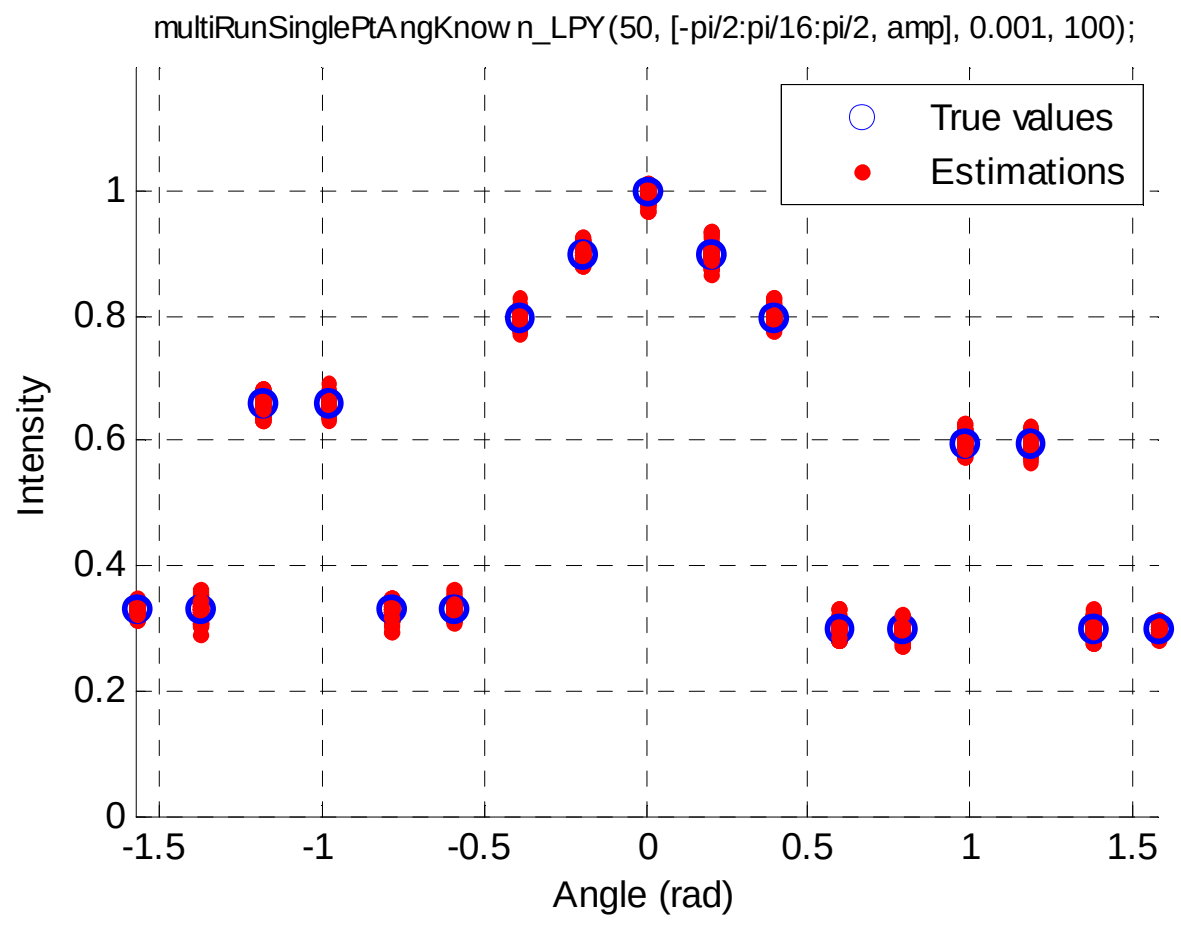

a)

Fig. 11. Distribution of estimates, by the “brainy” sensor with 100 elementary sensors, of intensities of a pattern formed by an array of 17 light sources with known directional angles. Standard deviation of sensor’s noise is 0.001. Standard deviation of intensity estimation is 0.002

## 8. Advantages and disadvantages of "brainy" light source sensors

As always, there is a tradeoff between good and bad features that have to enter into a system design. The designs discussed here have a number of advantages in common and a number of disadvantages as well.

The advantages include

- No precise optics needed. In some cases, no lens at all is necessary.
- The resolving power is not diffraction limited and depends only on the number of sensors and sensor's noise level
- The systems are inexpensive, compact, and rugged compared with imaging sensors.

The major drawback is extremely high computational complexity when good imaging properties for multiple sources are required

Such systems cannot give meaningful results if the a priori information does not apply. This is a simple case of the "No Free Lunch" theorem. What is ideal for a few point objects will certainly fail dramatically when applied to some other scenario. Whereas imaging systems do without a priori information injection and therefore do not have this problem.

The goal of this paper was to show multiple variations of our basic scheme of nonimaging data gathering to allow computational imaging with good results. We hope over time to explore many other aspects of these interesting devices.

## 9. Conclusion

If we know *a priori* that the scene is comprised of one or a few distant point sources, we can design non-imaging systems that gather the information needed to characterize the intensities and locations of those sources. All of these fall under the description "computational imaging." The directly recorded data do not comprise an image, but the appropriate mathematics can use the measurements to compute the image. We described several quite distinct approaches for accomplishing that. They all work well and can achieve better than diffraction limited performance. They are also smaller, more rugged, and less expensive than imaging systems.

The results we show prove the feasibility of "plane" sensors. They are in the good agreement with the above outlined theory for a single source and intuitive expectations for the multiple source cases. The feasibility demonstration contained here leaves many questions yet to be explored, for instance:

- Determination of the trade-off between the number of parameters of light sources, the number of "elementary" sensors and sensor's noise level
- Development of efficient computational algorithms for accelerating multi-dimensional optimization for ML-evaluation of multiple light source parameter, and, in particular, exploration of possible usage, for this goal, of parallel computational networks such as neural networks
- Determination of the trade-off between the number of parameters of light sources, the number of "elementary" sensors and sensor's noise level
- Determination, for given number of source parameters and given number of elementary sensors, of threshold levels of sensor's noise for which reliable parameter estimation is possible
- Efficient, tailorable designs

Possible extensions include

- Accurate light (and general) spectral analysis using multiple sensors with broad overlapping spectral sensitivities.
- Implementing this principle of optimal combination of multiple imperfect sensors in other metrology tasks
- Use of multiple units to obtain accurate 3D point source location.

**Acknowledgement**

L.P.Y thanks Chad Goerzen for the help in programming of the optimization procedures  HJC thanks William Culver for his insightful comments.